\documentclass{cimento}

\title{Ultra-strong Elelctric Fields and Vacuum Breakdown\ETC: 
Search for an Astrophysical Scenario}
\author{A.~Treves\from{ins:x},
R.~Turolla\from{ins:y}
S.B.~Popov\from{ins:z}}
\instlist{
\inst{ins:x} Department of Sciences, University of Insubria, via Lucini 3,
22100 Como, Italy
\inst{ins:y} Department of Physics, University of Padova, via Marzolo 8,
35131 Padova, Italy
\inst{ins:z} Sternberg Astronomical Institute, Universiteskii Pr. 13,
119899, Moscow, Russia}

\PACSes{\PACSit{00.00}{}
\PACSit{---.---}{\ldots}}
\begin{document}

\maketitle

\begin{abstract}
In some models of $\gamma$-ray bursts super-strong electric fields
($E\sim 10^{14} \ {\rm statvolt\, cm}^{-1}$) have been surmized. 
Such large fields may provide copious pair production through vacuum
polarization. Here we examine various astrophysical scenarios where huge
electric field can be present. It is shown that when the conditions for
quantistic vacuum breakdown are met, pair creation through 
conventional processes is likely to be always more important. Although of   
great physical interest, quantum pair production seems therefore of
little astrophysical relevance.
\end{abstract}

\section{Introduction}

Pair production by
quantum vacuum polarization (or vacuum breakdown, VB in the following) has
been
invoked by some authors (e.g. \cite{pre,li,r,ruf})
%Preparata et al. 1998, Lieu et al. 1999, Ruffini et al. 1999, Ruffini 1999)
in connection with the modeling of $\gamma$-ray bursts (GRBs).
However little attention was paid to the astrophysical picture where
the VB condition might occur.

Here we examine various astrophysical scenarios where large electric field
may be expected and explore the possibility of inducing the VB.
In section 2 large B-field pulsars (``magnetars'') are considered,
following a proposal by Usov \cite{us}. In section 3
the case of a magnetic field frozen in the matter accreting onto
black
hole is examined, and in section 4 we consider the E-field due to charge
separation, induced by a flash of radiation propagating through a plasma,
in the context of the GRB scenario proposed by Lieu et al \cite{li}. 

Since we are interested in a preliminary assessment of the relevance of
quantum pair
production under astrophysical conditions, we assume that the condition for 
vacuum breakdown (Klein instability) 
is given by the semi-classical expression (see e.g. \cite{no})
\begin{equation}
\label{break}
eE\lambda_c=2m_ec^2
\end{equation}
where $\lambda_c$ is the Compton wavelength of the electron, which gives

\begin{equation}
\label{break1}
 E> 2m_ec^2/e\lambda_c = E_c\approx2\times 10^{14} \ {\rm statvolt\, cm}^{-1}.
\end{equation}

\section{Highly magnetized neutron stars}

Huge electric fields are produced in pulsars by the rotating 
magnetic field entrapped in the neutron star crust.

As far as we know, the possible appearance of VB was first proposed in a
remarkable paper by Usov \cite{us}. It was suggested that GRB should be the
progenitors of ultra-high magnetic field pulsars (or magnetars),
proposed at  about the same time by Duncan and Thompson \cite{dun} in 
connection with soft gamma repeaters. Usov made the rather bold, at that 
time, assumption that GRBs are at cosmological distances, and took as 
characteristic parameters for the newly
formed magnetar a period $P=0.5$ ms and a B-field $B=3\times 10^{15}$ G. 
Some years later magnetars were actually discovered and the main parameters
of the prototype of this class of sources,
SGR 1806-20 \cite{ko}, are $P=7.5$ s, $\dot 
P=2.6\times 10^{-3}  {\rm s/yr}^{-1}$, yielding $B=8 \times 10^{14}$ G, in 
remarkable agreement with the predictions by Usov.

Following Usov, we take for the electric field the expression
of Goldreich and Julian \cite{gold}, which at the neutron star surface reads

\begin{equation}
\label{erot}
% E\sim l\omega B/c
E\sim R\omega B/c
\end{equation}
where 
%$l\sim \theta_c R$, $\theta_c$ is ??? and 
$R\sim 10^6$ cm is the 
neutron star radius. Taking as reference values for the magnetic field and
the period $10^{15}$ G and 1 ms respectively (note that these values are
a little bit less extreme than those used by Usov), 
the condition for pair creation (eq. [\ref{break1}]) becomes

\begin{equation}
% B_{14}P_2^{-3/2}>2,
B_{15}P_1^{-1}>1
\end{equation}
which formally may be satisfied in an object like SRG 1806-20
if it was born with a short period.
However, there are good reasons to suspect that $E$ can never
reach  $E_c$, because competing processes tend to damp the field. 
Pair production by photon splitting in a magnetic field
(the $1-\gamma$ pair production) is expected to be extremely effective in
magnetars. This process was first
considered by Sturrock \cite{st} in the context of pulsar electrodynamics 
and has a threshold

\begin{equation} 
 B_pE_{\gamma}\sim 4\times 10^{18} \ {\rm eV\,G}
\end{equation}
where $E_{\gamma}$ is the photon energy and $B_p$ is the perpendicular
component of the magnetic field. So, $\gamma$-rays of energies
about $10^7$ eV moving in magnetic field $\sim 10^{12}$ G will produce
$e^+-e^-$ pairs. The pair cascade is sustained by synchrotron
(curvature) photons and $1-\gamma$ pair production. This in Sturrock's
scenario produces sparks in the electroactive gap. Though the process in a
way is a sort of vacuum breakdown, it is largely mediated by the magnetic
field. The sparking inhibits the growth of the $E$-field well below $E_c$.
We note, however, that recent observations of highly magnetized radio pulsars
with $B >4\times 10^{13}$ G \cite{cam},
support the idea of Usov and Melrose \cite{um}, that
photon splitting is inhibited by polarization selection rules, so 
the idea that 
$1-\gamma$ process are more important than vacuum breakdown should be taken
with care.

Another astrophysical scenario, which can be relevant for the present
discussion, is the coalescence of binary neutron stars. This is nowdays
a popular scenario for producing GRBs at cosmological distances and was 
first suggested by Blinnikov et al. \cite{bl}.
In this case the electric field is generated by the fast orbital motion of
the magnetized neutron stars as they spiral in, much in the same way as 
the rotating dipole produce the $E$-field in an isolated neutron
star (see eq. [\ref{erot}]). This possibility was considered by 
Vietri \cite{ve} (see also \cite{lip}), 
who found that large electric fields may indeed form

\begin{equation}
E\sim \frac{v}{c} B \sim 5\cdot 10^{14}\, 
\left(\frac{R}{10^6\, {\rm cm}}\right)^{-7/2}\left(\frac{B}{10^{15}\, 
{\rm G}}\right) \ {\rm statvolt}\, {\rm cm}^{-1}\,. 
\end{equation}
where the two neutron stars are assumed to be nearly in contact (typical
separation $\sim R$, the star radius) and moving at Keplerian velocities.

Ruffini and Treves \cite{rut} proposed that a neutron star, rotating
in vacuo should be endowed with a net charge, as it follows from a variational
principle. For short periods the electric field has a form very
similar to that of eq. ({\ref{erot}). If this is the case, the same
considerations on VB presented earlier in this section  
should apply.

High electric fields can be also generated when the inner edge of a
Keplerian accretion disk rotates faster than the neutron star \cite{ch}. 
Also in this case $E\sim B$ and all the above discussion 
is again valid, but in realistic astrophysical situations, like X-ray binaries,
the field turns out $\sim 10^9$ statvolt cm$^{-1}$, too low for producing VB.

\section{Accreting Black Holes}

 We assume, as in the case of pulsars, that the electric field is the
Lorentz transformation of the $B$-field. However it is much harder to associate
a huge $B$-field to a black hole (BH) than to a neutron star.

A rather sound scenario is that of taking an accreting BH and considering
the compression of the $B$-field entangled in the infalling material.
Accretion compresses the $B$-field, and if the flux is conserved, the field is
amplified. Here we give a quasi-Newtonian description and take spherical
symmetry (see e.g. \cite{tma}).
The maximum field is generally taken to be of the equipartition value, 

\begin{equation}
B\sim U_G\sim \frac{\dot M}{4\pi r^2 v}\frac{GM}{r}\,,
\end{equation}
where $v$ is the infall velocity and $\dot M$ the accretion rate. 
At $r\sim r_G=2GM/c^2$,  $v\sim c$ and one has

\begin{equation}
 B\sim \left(\frac{2\dot M c}{r_G^2}\right)^{1/2}\sim 5\times  10^8 
\xi^{1/2} \left(\frac{M}{M_{\odot}}\right)^{-1/2}
\end{equation}
where $\xi$ is the accretion rate in units of the Eddington rate 
$\dot M_E=4 \pi G m_p M/\sigma_T c $.
Assuming again $E\sim B$ the VB condition becomes

\begin{equation}
\label{accrvb}
 \xi >2\times 10^{11}\frac{M}{M_{\odot}}.
\end{equation}
In steady state $\xi$ is essentially the inverse
of the efficiency. Even in the case of a solar mass BH, accretion should
be $\sim 10^{12}$ larger than the Eddington rate in order to enter the VB 
regime. Note that eq. (\ref{accrvb}) indicates that the more massive the BH 
is,  the more difficult is to fulfill eq. (\ref{break1}).
Suppose that  $\xi$ could be taken arbitrarily large. Very high accretion 
rates can be reached if neutrino losses dominate: the  
``Eddington'' rate for neutrino losses is infact much
larger than that associated to photons. 
Still, as in the case of pulsars, since the electric field
is due to a huge $B$-field, synchrotron-$1-\gamma$ pair production
showers will quench the $E$-field below the critical value.

\section{Charge separation by a flash of radiation}

Here we focus on the $E$-field produced by the charge separation induced on a 
globally neutral plasma by a strong radiation field.
The basic idea is that the radiative force on 
electrons is a factor $\sim (m_e/m_p)^2$, the ratio of the Thomson 
cross-sections, larger than on ions. Schwartsman \cite{sh} 
considered a spherical accretion model, showing that in order that the
electron and ion to fall in at the same speed, the central body must acquire
a positive charge. Michel \cite{mi} extended the treatment to the case of an 
accreting BH. Maraschi et al. \cite{mar} 
studied the flow and the $E$-field for luminosities approaching the
Eddington limit, $L_E$. More recently this problem was reconsidered by 
Turolla et al \cite{ttz}, 
who in particular examined its efficiency to accelerate positrons 
(the Schwartsman accelerator).
In \cite{tt} it was discussed under which condition the 
$E$-field may reach the
breakdown value within stationary models. Supposing that $L\sim L_E$ the only
free parameter is the mass of the accreting BH.
VB is possible only for $M\le 10^{20}$ g, a value which makes
dubious the astrophysical relevance of the mechanism.

The main restriction of the previous treatment is the stationarity
hypothesis, which limits the luminosity to $L_E$. Here we
overcome this restriction, and explore the conditions for which a burst of
radiation of arbitrary luminosity may induce an $E$-field at the VB level.

Consider a spherical shell of hydrogen plasma, and let it be transparent to
radiation, which is characterized by luminosity $L$. Let the
electron-photon cross section be the Thomson one,
$\sigma_T=0.66\times 10^{-24}$ cm$^{2}$. The charge separation induced by the
radiation produces, at distance $r$, an electric field of intensity 
\cite{sh,ttz}: 

\begin{equation}
\label{Echsep}
 E=\frac{\sigma_T L}{4\pi cer^2}.
\end{equation} 
Note that  eq. (\ref{Echsep}) should be regarded only as an estimate
of the electric field, since a non-stationary situation is considered.
Note also that in the present picture the electric field is not in vacuo, but 
in a plasma, charge-separated by the photons. As an order of magnitude we still
take equation (\ref{break}) as valid to estimate the VB condition.

Combining equations (\ref{break}) and (\ref{Echsep})
we obtain the critical luminosity which should
induce VB

\begin{equation}
 L_c=\frac{8\pi r^2 m_e c^3}{\sigma_T \lambda_c}.
\end{equation}
Having in mind models of GRBs, we scale the size of the emiting region 
to $10^6$ cm, so the above equation becomes

\begin{equation}
 L_c= 4\times 10^{51} r_6^2 {\rm erg\, s}^{-1}.
\end{equation}   
This typical value of the luminosity is of the same order of that of GRBs.

By introducing the ``vacuum breakdown'' parameter

\begin{equation}
l_c=\frac{L}{L_c}=\frac{\sigma_T\lambda_cL}{8\pi r^2 m_e c^3}
\end{equation} 
it is apparent that $l_c$ is related to the well-known 
``compactness parameter'', $l=L\sigma _T/4\pi r m_e c^2$
\cite{cav}, which measures the opacity of a source to
photon-photon interaction. In fact one has

\begin{equation}
\label{lcandl}
 l_c=2\frac{\lambda_c}{r}l\sim 5\times 10^{-16}l\, r_6^{-1}
\end{equation}
which implies that s source in ``breakdown'' condition (i.e. with $l_c\sim 1$)
is also extremely compact, $l\gg 1$.
A lower limit to the photon energy density under VB conditions is simply

\begin{equation}
 U=\frac{L_c}{4\pi r^2 c} \sim \frac{2 m_ec^3}{\sigma_T \lambda_c}
\sim 10^{28} {\rm erg\, cm}^{-3}.
\end{equation}
If the photon spectrum is a blackbody, one can obtain a lower limit
to the typical photon energy

\begin{equation}
\label{ephot}
 E_\gamma\sim (U/a)^{1/4}\sim 3 {\rm MeV}.
\end{equation}
Note that this limit is independent of the source size $r$.

\section{Conclusions}

Present models of GRBs indicate that photon fluxes leading to the VB
condition $l_c\ge 1$ may be achieved. This corresponds to a huge
compactness parameter (see eq. [\ref{lcandl}]). If the spectrum is not 
far from a blackbody, equation (\ref{ephot})
shows that photon-photon interaction will produce a pletora of pairs.

Although in principle the photon energies may be less than the value
implied by eq. (\ref{ephot}), remaining below the threshold of pair creation, 
the situation may be considered in a ``gedanken experiment'', but it seems 
to us very unlikely in a realistic astrophysical situation.

Our conclusion is that in GRBs vacuum polarization by electric field
produced by an extreme luminosity may occur, but its  consequences will be
masked by the multitide of pairs produced by photon-photon interaction.

\end{document}